\documentstyle[ichep,12pt]{article}
\def\overtext#1{$\overline{#1}$}

\def\eg{{\it e.g.}}
\def\ie{{\it i.e.}}
\def\VEV#1{\left<#1\right>}

\newif\ifproceeding
\proceedingtrue
\title
{Stringy Origin of
Neutrino Masses within
the Minimal Supersymmetric Standard Model\ifproceeding\thanks{Invited
Talk Presented at the XXVI
 International Conference on High Energy Physics, August 6--12,
1992, Dallas, Texas.}\fi}
\author{Mirjam Cveti\v c\ifproceeding\thanks{email
cvetic@cvetic.hep.upenn.edu}\fi \\ Department of Physics\\
University of Pennsylvania\\
Philadelphia, PA 19104--6396\\}

\begin{document}
\finalcopy
\ifproceeding\pagestyle{myheadings}
\markright{\vbox{\rm\noindent UPR--528--T\hfill\vskip 3mm\noindent September,
1992\hfill\vskip4mm}}\fi
\maketitle
\abstract{
We present a ``gravity-induced'' seesaw mechanism,
which  accommodates neutrino masses
($m_\nu\propto
m_u^2/M_I$, with $m_u$ the corresponding
quark mass and
$M_I\simeq 4\times10^{11}$ GeV)
compatible with the MSW study of the Solar neutrino deficit within
the minimal supersymmetric Standard Model
(the grand desert with the
gauge coupling unification
at $M_U\simeq
2\times10^{16}$ GeV).
We show  that  for
 large radius ($R^2/\alpha '={\cal O}(20)$)
 Calabi-Yau spaces, threshold corrections
ensure
 $M_U^2=M_C^2/{\cal O}
(2R^2/\alpha')$ and the magnitude of the non-renormalizable
terms in the superpotential yields
$M_I=
{\cal O}(e^{-R^2/\alpha'})M_C$. Here
 $M_C=g\times 5.2\times
 10^{17}$GeV
 is the
  scale of the tree level (genus zero)
 gauge coupling ($g$) unification.
}

\vskip-1pc
Precise
data from the LEP experiments
indicate that the gauge couplings of the Standard Model meet at
$M_U\simeq (1-4)\times
10^{16}$ GeV in the minimal supersymmetric extension of the
Standard Model.\cite{gauge}
  Another set of intriguing data arise from the Solar
neutrino experiments.  The deficit of Solar neutrinos can most
efficiently be explained through the MSW\cite{MSW}
mechanism of matter-enhanced neutrino
oscillations.  In particular, current data favor\cite{BKL}
the mass splitting of
the electron and muon neutrinos to be
$\Delta m^2 \equiv
|m_{\nu_\mu}^2-m_{\nu_e}^2|$  in the range
$ (1-16) \times 10^{-7}  {\hbox{eV}}^2$.
 In     the grand unified gauge theories
(GUT's) the
seesaw scenario\cite{GellM}
the light neutrino masses are given by:
$m_{\nu_{e,\mu}}\simeq c\ m^2_{u, c}/M_I$, where $m_{u,c}
$ are the corresponding  quark masses
and $c\simeq 0.05-0.09$ is
a factor due to the renormalization down to the low energy
scale\cite{BKL}.
This implies that\ifproceeding\newpage\fi
$M_I\simeq
(4 \pm 3) \times10^{11}$ GeV, the central value corresponding to
the neutrino mixing angle
$\theta_{\nu_\mu \nu_e}
\sim \theta_{Cabibbo}
$.

  Each of the two sets of experimental
data has therefore an elegant theoretical explanation.
 Unfortunately, the two
theoretical models are mutually exclusive at first glance.
Here, we  present a scenario\cite{CL}
which implements  the neutrino masses in the minimal
supersymmetric  Standard Model in such a way that there
is still a grand desert with the
gauge coupling unification
at $M_U\sim (1-4) \times 10^{16}$GeV, while the
effective scale $M_I$ governing the neutrino masses
is in the range of  $(4 \pm 3) \times 10^{11}$ GeV.
We are proposing  a ``gravity-induced''\cite{BEG,ABS}
seesaw mechanism (an extension of a mechanism proposed by
Nandi and
Sarkar\cite{Nandi} within the context Calabi-Yau vacua with
$E_6$ gauge group\cite{MV}),
realized through
an interplay between the nonrenormalizable and renormalizable
terms in the superpotential,
as the origin of the neutrino  masses.

The essence of the idea is based on a supersymmetric
theory with an
extended gauge symmetry, which
contains
an additional sterile
neutrino (a Standard Model singlet), and a {\it
restricted}  representation of the Higgs fields. Such
fields
break the extended gauge symmetry
at the scale $M_U$. However, they cannot
give the sterile neutrino
a large Majorana mass  proportional to $M_U$ through
  the renormalizable (cubic) terms  in the
superpotential.  On the other
hand, through
the nonrenormalizable (\eg  , quartic)  terms
of the superpotential, which
are  suppressed by a scale  $M_{NR}>M_U$,
such Higgs fields can give a
 Majorana mass of order $M_U^2/M_{NR}<M_U$.

One can demonstrate\cite{CL}
  the  gravity-induced seesaw
in an explicit (minimal)
model with  all the
essential features, by choosing
an enhanced
gauge symmetry,  $SU(3)_C\times
SU(2)_L\times U(1)_{Y}\times U(1)_{Y'}$, where $Y$ is the
ordinary weak hypercharge.
The matter
consists of the particle content of
the minimal Standard Model
as  well as of the
Standard Model singlets,
 $ L_i$, $S_1$ and $\bar S_1$.
$L_i$ supermultiplets with $i$ = (1,2,3)
contain  a sterile neutrino which accompanies
each of the three families.
 $S_1$ and $\bar S_1$ contain the Higgs fields
which break the enhanced gauge symmetry with
VEV's  of order $M_U$.

Consistent with the anomaly constraint
we choose the  following
 values for the $Y'$ charges:
quark  $SU(2)_L$ doublets,
$u_L^c$ quarks and  $e_L^c$  leptons
have (-1),
lepton   doublets  and
$d_L^c$ quarks
have (+3),  $L_i$ and $S_1$ have (-5),  $\bar S_1$ has (+5), while
Higgs doublets $H_{(1,2)}$ have (-2) and  (+2),
respectively.\cite{footII}
In the neutrino sector,
the only renormalizable terms
allowed in the superpotential are of the
type $W=L_i\nu_i H_2$ which yields
the neutrino mass matrix:
\begin{equation}
\left[\matrix{0&m\cr
             m&0\cr}\right],\label{threem}
\end{equation}
where $m$ is proportional  to the  $VEV$ of the
Higgs doublet $H_2$. Since $H_2$ gives mass
to the quarks as well,
$m$ is of the order of the corresponding quark masses.
On the other hand, the only allowed  nonrenormalizable term
in the superpotential with a leading contribution to the neutrino
mass matrix is of the type  $ W_{NR}=
L_iL_i\bar S_1\bar S_1/M_{NR}$.
This              modifies the neutrino mass matrix:
\begin{equation}
\left[\matrix{
              0&m\cr
              m&M_I\cr}\right],\label{nrm}
\end{equation}
where $M_I=M_U^2/M_{NR}$.
            The
quantum numbers prevent  the contribution of any
non-renormalizable term to the
 $\nu \nu$ and $\nu L$
 masses  that would be of the
 order of  $M_U^K/M_{NR}^{K-1}$ for any $K>   1$.

While the above scenario is appealing on its own terms,
 its origin can be motivated
 from the properties of superstring
vacua, which  provide
 a natural framework for  the
restricted representation of the chiral
supermultiplets.
This can be exhibited in explicitly\cite{Nandi,CL}
in the case  of $E_6$ gauge group and the restricted
representation (${\bf 27}$'s  and $\overline{{\bf 27}}$'s
of $E_6$) of the matter supermultiplets.

However, the requirement of the
minimal supersymmetric Standard Model particle content
below $M_U
\simeq (1-4)\times 10^{16}$ GeV  imposes  severe constraints
on the  allowed superstring vacua.
$(2,2)$ string vacua, {\it
e.g.,} Calabi-Yau manifolds with gauge and spin connection identified,
possess a large number of additional multiplets.
In particular, for vacua without Wilson lines, the gauge
group is
$E_6$, with {\bf27}'s, ${\overline{\bf27}}$'s, and {\bf1}'s of $E_6$.
Some of the particles in these multiplets
acquire large masses if there are flat directions in
the space of specific string vacua.
Finding flat
 directions
allows one to give large
VEV's  to fields in a
particular set of {\bf27}'s and ${\overline{\bf27}}$'s, which in
turn   can
give mass to some of the
unwanted massless multiplets. At the  same time,
 $E_6$ is broken down to
$SO(10)$ or $SU(5)$.
Such directions were found for
orbifolds\cite{Cvetic,Font} as well as for a class of Calabi-Yau
manifolds\cite{Greene} based on Gepner's\cite{Gepner}
 construction.

In addition,
Wilson lines
allow for  a breakdown of the simple
gauge group ($E_6$, $SO(10)$, or $SU(5)$) to a direct product of
simple groups and $U(1)$'s.
It is in general
possible\cite{FontII,Greene}
to introduce Wilson lines which  break the gauge group down to the
Standard Model. At the same time, this procedure decouples
a large number of unwanted modes.

 Thus, a viable scenario
is to construct (2,2) string vacua with  flat directions
as well as  Wilson lines.
However,
in spite of the progress made  in the construction of such string
vacua there  exists no explicit
example of      a supersymmetric string vacuum which would contain
only the minimal Standard Model particle spectrum below the gauge
coupling unification scale.
On the other hand,
string theory
can   shed light on the scale of the gauge coupling unification $M_U$,
and  the  magnitude  the nonrenormalizable
terms in the superpotential in a quantitive manner.

While string vacua in general
do not possess gauge group grand unification, there is a  notion
gauge coupling unification. The
scale $M_C$ associated with the gauge coupling unification
at the tree level of the string theory is
determined\cite{Kaplunovsky} in the ${\overline {DR}}$ scheme by
the value of the Planck mass $M_{Pl}$
and of the gauge coupling $g$
in the following way:
\begin{eqnarray}
M_C&=&{e^{(1-\gamma)/2}\sqrt2\over{3^{3/4}\sqrt{\pi\alpha '}
}}=g\times {e^{(1-\gamma)/2}\over{3^{3/4}4\pi}} M_{Pl}\nonumber\\
&=&  g\times 0.043M_{Pl}\nonumber\\
 &=&  g \times 5.2\times10^{17}
\hbox{~GeV}.\label{mc}
\end{eqnarray}
where $\gamma=0.57722$ is the Euler constant,
$g^2=32\pi/(\alpha 'M_{Pl}^2)$, with $g$ defined according to the $GUT$
convention               and $M_{Pl}=1.2\times 10^{19}$GeV.
For the expected value  $g \sim 0.7$
this is one order of magnitude too large
compared with
$M_U\sim(1-4)\times
10^{16}$ GeV, which is the scale of the
gauge coupling unification  of the minimal supersymmetric Standard Model.
However, threshold effects, \ie, genus one corrections to the gauge
couplings, can split the gauge couplings at
$M_C$, thus in principle allowing for an effective unification scale
$M_U<M_C$.
The part of the threshold corrections
which does not depend on the vacuum expectation values (VEV's)
of fields is small and at most few percent (\ie\
${\cal O}(1)/16\pi^2$) correction to the tree level gauge
coupling.\cite{Kaplunovsky}
  This  result is to    be expected;
heavy modes should not affect the evolution of gauge couplings at
scales  much  lower than
$M_{Pl}$.\cite{Weinberg}
 On the other hand, field dependent threshold
corrections  could be large; namely, massless fields which could
acquire VEV's (\ie\ of the order of $M_U$) can drastically affect
the threshold corrections.

In  (2,2)  string vacua,
 there exist
 massless chiral superfields,  moduli ($T_i$)
which are  in one to
one correspondence with the matter fields, {\it i. e.},
${\bf 27}$'s and $\overline{\bf 27}$'s of $E_6$.
Since moduli do not have a potential to all orders  in string loops,
their VEV's
parametrize a whole class of string vacua.
  Explicit calculations of the moduli dependence
for the threshold
corrections for a class
of orbifolds are given in Refs. \cite{Kaplunovsky,Louis}.
Extensive study\cite{Ross} of  threshold corrections in orbifolds
indicate that $M_U<M_C$ if the massless spectrum possesses certain
modular weights which should   also be
 compatible with the
target space one-loop modular anomaly.
In such examples one would obtain $M_U={\cal O}(e^{(-cR^2/\alpha ')})
M_C$ when the orbifold radius  is large
($R^2/\alpha '\gg 1$ ).
The positive coefficient $c$ depends\cite{Ross}  on the modular
weights of the massless states.
As  we shall see later, the heavy Majorana
mass turns out to be
$M_I={\cal O}(e^{-c'R^2/\alpha ')})M_C$.
In order to ensure $M_U\sim 10^{16}$ GeV and
$M_I\sim 10^{12}$GeV , this in turn involves  detailed
constraints on coefficients   $c$ and $c'$.

In the following, we shall pursue a different approach, \ie ,
study of smooth Calabi-Yau spaces.
 Calabi-Yau spaces are related to the
underlying solvable
$(2,2)$ conformal field theory (CFT) (\eg\
orbifolds,\cite{DHVW}
 and
Gepner's models\cite{Gepner}) by
taking the VEV's of {\it
all}
 the moduli $\VEV{T_i}>>1$.
The internal space  is then
large and smooth everywhere and one can reliably use
  ({\it e.g.,} blown-up orbifolds\cite{CvetMary})
 the $1/\VEV{T_i}$ expansion of the point field limit
approach, thus obtaining qualitatively different
results.\cite{Dixon,MSS}
A large Calabi-Yau  space
 asymptotically approaches
     the untwisted sector of the
orbifolds,  \ie\ it is a smooth,
 almost flat space.
Thus, the massive modes (Kaluza-Klein states)
contributing to the threshold corrections
are those of $N=4$ sector (for each vector supermultiplet there are
three chiral superfields) and therefore
 their contributions go to zero
with the inverse  powers of the moduli VEV's.\cite{Dixon}
The leading  threshold  corrections  therefore arise only from the
light fields and they can be calculated using  a field
theory one-loop calculation.\cite{LouisII}
  Such corrections
 are milder, logarithmic in
the VEV's of moduli.

The fields theory  calculation for the (2,2) vacua without an enhanced
gauge symmetry yields\cite{DF}
 the following dependence
on the $T_i$ moduli associated with the scaling deformations
((1,1) forms)):
\begin{eqnarray}
&&\Delta\left(
{{16 \pi^2}\over{g_{E_6}^2}}\right)
={1\over3}b_{E_6}K_1\nonumber\\
&&+T(27)
\left[2\log\det\left(
{{\partial^2K_1}\over{\partial T_a\partial
T_b^\ast}}\right)
-{4\over3}b_{(1,1)}K_1\right].\nonumber\\
&&\hbox{\hfill}\label{fttc}
\end{eqnarray}
Here  $b_{E_6}=C_{E_6}-b_{(1,1)}T(27)-b_{(1,2)}T(\hbox{\overtext{27}})$.
is related to the  one-loop   $N=1$ beta function $\beta_{E_{6}}=
b_{E_{6}}g^3/16\pi$
with $C_{E_6}$ corresponding to the
    quadratic Casimir operator  of $E_6$ gauge group and
$T(27),\ T(\overline{27})$ are
$tr(Q_{(27,\overline{27})}^2)$, respectively.
Note, that there  are $b_{(1,1)}\ (b_{(1,2)})$
$\bf{27}$'s ( $\overline{\bf{27}}$'s), in one to one correspondence
with the number of moduli associated with the scaling deformations
(complex structure deformations).
The K\" ahler potential for the moduli  $T_i\gg 1$ is of the form
 $K_1=\log\sum_{i,j,k=1}^{b_{(1,1)}}\left(T_i+T_i^\ast\right)
\left(T_j+T_j^\ast\right)\left(T_k+T_k^\ast\right)h_{ijk}$
where $h_{ijk}$ are intersection numbers of $(1,1)$ forms.
Taking
 $\VEV{T_i}
= \alpha_iT        $ with $T\gg1$ corresponding to an overall
size of the manifold  and
$\alpha _i={\cal O}(1)$ one finds:\cite{Dixon,MSS}
\begin{equation}
 \Delta\left(
  {{16\pi^2}\over{g_{E_{6,8}}^2}}\right)
=-b_{E_{6,8}}[\log\left(T+T^\ast\right)+{\cal O}(\log\alpha_i)]
\label{esix}
 \end{equation}
which are proportional to the
  $N=1$ beta function $\beta_{E_{(6,8)}}=
b_{E_{(6,8)}}
g^3/16\pi$
 of $E_6$ ($E_8$) gauge groups.
This implies that  while
the slope of the running gauge couplings is not
changed, the effective
gauge coupling  unification scale is
lowered:
\begin{equation}
M_U^2={{M_C^2}\over{(T+T^*)}}=
{\cal O}({{M_C^2}\over{2R^2/\alpha'}})\label{mu}
\end{equation}
The above results apply  only to simply connected
Calabi-Yau spaces. The gauge group   $E_6$
can be broken if Wilson lines are introduced. The  study of threshold
corrections in this case is under way.\cite{ME}
We proceed under the assumption that
the  same features persist also in this case.

 From eq. (\ref{mu}) one then sees that for
$R^2/\alpha'\sim{\cal O}(20)$
the gauge unification scale is lowered
to $M_U \sim 6\times
10^{16}$ GeV, which is slightly too large. However,
eq. (\ref{mu}) relates $M_U^2$  to  $R^2/\alpha '$
only by  orders of magnitude. Thus,
an additional factor
of $2$  in the relation of an overall modulus $Re
T$ to $R^2/\alpha '$
 enables  one to obtain $M_U$ in the preferred
range $4\times 10^{16}$ GeV.

We turn now
address the size of the nonrenormalizable terms.
 In
string theory the magnitude of the coefficient
 $M_{NR}$ is proportional to $M_C$.  However, one can
prove explicitly\cite{Lee}
that for all  $(0,2)$ string vacua
the nonrenormalizable terms are suppressed by an additional
factor $e^{-R^2/\alpha'}$, {\it i.e.,} the origin of the
nonrenormalizable terms is due only to world-sheet
instanton effects.
This is a
 stringy result, proven
explicitly on (blown-up) orbifolds\cite{Dine}
as well as
in sigma model
 perturbations\cite{Lee} of Calabi-Yau manifolds.
 Therefore:
\begin{equation}
{1\over M_{NR}}={{{\cal O}(e^{-R^2/\alpha'})}\over M_C}.
\label{mnr}
\end{equation}
By choosing vacuum expectation values along the flat direction
to be $M_C$ (the only
 natural scale in the four-dimensional string vacuum)
 nonrenormalizable terms
 yield the heavy Majorana mass:
\begin{equation}
M_I={M_C^2 \over M_{NR}}={\cal O}(e^{-R^2/\alpha'})M_C.\label{mi}
\end{equation}
It follows from
eq. (\ref{mu}) that
we need $R^2/\alpha'={\cal O}(20)$ in order to achieve $M_U\sim10^{16}$
GeV. In this case
$M_I\sim10^{-8}M_C\sim10^{10}\hbox{~GeV}$.
Although these are only order of magnitude statements, it is
instructive to set the coefficients in eqs. (\ref{mu}) and (\ref{mi})
equal to unity. In that case the range $M_I \sim (4 \pm 3)
\times 10^{11}$ GeV suggested by the Solar neutrino deficit implies
$R^2/\alpha' \sim (13 - 15)$, yielding a slightly too large
$M_U \sim 7 \times 10^{16}$ GeV.

To summarize, the desired scale of the gauge coupling unification
 $M_U\sim 2 \times 10^{16}$ GeV and the scale of Majorana
neutrino masses
$M_I\sim4\times
10^{11}$ GeV, may be achieved
in a class of large radius ($R^2/\alpha'={\cal O}(20)$)
 Calabi-Yau spaces
 allowing
for $M_U^2=M_C^2/{\cal O}(2R^2/\alpha')$ and $M_I={\cal
O}(e^{-R^2/\alpha'})M_C$.

A major part of the presented work has been done in collaboration with
P. Langacker. This research was  supported in part by the
U.S. DOE Grant DE-AC02-76-ERO-3071,
and by a junior faculty
SSC fellowship (M.C.).


\begin{thebibliography}{99}
\bibitem{footII}The choice of the particle
spectrum and the $U(1)_{Y'}$   quantum numbers are motivated  by
$SO(10)$ gauge symmetry.
\end{thebibliography}
\end{document}